\documentclass[useAMS,usenatbib]{mn2e}

\usepackage{epsfig}
\usepackage{amsmath}
\usepackage{amssymb}
\usepackage{ifthen}
\usepackage{txfonts}
\usepackage{rotating}
\usepackage{url}
\usepackage{varioref}
\usepackage{verbatim}
\usepackage{latexsym}
\usepackage{graphicx}

\DeclareSymbolFont{cmletters}{OML}{cmm}{m}{it}
\DeclareMathSymbol{v}{\mathalpha}{cmletters}{"76}

\usepackage{multirow}



\DeclareSymbolFont{cmletters}{OML}{cmm}{m}{it}

\DeclareMathSymbol{v}{\mathord}{cmletters}{"76}

\def\be{\begin{equation}}
\def\ee{\end{equation}}

\newcommand{\bR}{{\bf{R}}}

\newcommand{\Bvec}{{\underline{B}}}

\newcommand{\erg}{{\rm\,erg}}

\newcommand{\cut}[1]{\hbox{}}

\DeclareSymbolFont{cmletters}{OML}{cmm}{m}{it}
\DeclareMathSymbol{v}{\mathalpha}{cmletters}{"76}

\usepackage{subfigure}
\usepackage{ifthen}
\usepackage[usenames,dvipsnames]{color}
\usepackage{graphicx}

\defcitealias{bz77}{BZ77}

\newcommand{\eff}{{\eta}}

\newcommand{\rhorest}{{\rho}} % rest-mass density
\newcommand{\ug}{{e_{\rm gas}}} % gas internal energy density
\title[Efficiency of Super-Eddington MADs]
{Efficiency of Super-Eddington Magnetically-Arrested Accretion}

\author[J.~C.~McKinney,
L.~Dai,
M.~Avara]
{Jonathan C. McKinney$^1$\thanks{\hbox{E-mail: jcm@umd.edu~(JCM)}},
Lixin Dai$^1$,
Mark Avara$^2$,
\\
 $^1$University of Maryland at College Park, Dept. of Physics, Joint Space-Science Institute, 3114 Physical Sciences Complex, College Park, MD 20742, USA \\
 $^2$University of Maryland at College Park, Dept. of Astronomy, CSS 1231, College Park, MD 20742, USA \\
  }{
}

\begin{document}
\date{Accepted 2015.  Received 2015; in original form 2015.}
\pagerange{\pageref{firstpage}--\pageref{lastpage}} \pubyear{2015}
\maketitle

\newcommand{\bea}{\begin{eqnarray}}
\newcommand{\eea}{\end{eqnarray}}
\newcommand{\pdder}[2]{\frac{\partial^2 #1}{\partial#2^2}}
\newcommand{\pder}[2]{\frac{\partial#1}{\partial#2}}
\newcommand{\der}[2]{\frac{{\rm d}#1}{{\rm d}#2}}
\newcommand{\derln}[2]{\ensuremath{\frac{{\rm d\,ln}\, #1}{{\rm d\,ln}\, #2}}}
\newcommand{\pderln}[2]{\ensuremath{\frac{\partial\,\rm ln\,#1}{\partial\,\rm ln\,#2}}}

\newcommand{\AS}[1]{\textbf{\color{Magenta}#1}}
\newcommand{\RN}[1]{\textbf{\color{Green}#1}}
\newcommand{\AT}[1]{\textbf{\color{Blue}#1}}
\newcommand{\YZ}[1]{\textbf{\color{Orange}#1}}
\newcommand{\koral}{\texttt{koral}}
\newcommand{\harmrad}{\texttt{harmrad}}
\newcommand{\harm}{\texttt{harm}}

\def\bE{\bar{E}}
\def\bR{\bar{R}}
\def\bu{\bar{u}}

\label{firstpage}

\begin{abstract}

The radiative efficiency of super-Eddington accreting black holes
(BHs) is explored for magnetically-arrested disks (MADs), where
magnetic flux builds-up to saturation near the BH.  Our
three-dimensional general relativistic radiation magnetohydrodynamic
(GRRMHD) simulation of a spinning BH (spin $a/M=0.8$) accreting at
$\sim 50$ times Eddington shows a total efficiency $\sim 50\%$ when
time-averaged and total efficiency $\gtrsim 100\%$ in
moments. Magnetic compression by the magnetic flux near the rotating
BH leads to a thin disk, whose radiation escapes via advection by a
magnetized wind and via transport through a low-density channel
created by a Blandford-Znajek (BZ) jet.  The BZ efficiency is
sub-optimal due to inertial loading of field lines by optically thick
radiation, leading to BZ efficiency $\sim 40\%$ on the horizon and BZ
efficiency $\sim 5\%$ by $r\sim 400r_g$ (gravitational radii) via
absorption by the wind.  Importantly, radiation escapes at $r\sim
400r_g$ with efficiency $\eta\approx 15\%$ (luminosity $L\sim 50L_{\rm
  Edd}$), similar to $\eta\approx 12\%$ for a Novikov-Thorne thin disk
and beyond $\eta\lesssim 1\%$ seen in prior GRRMHD simulations or slim
disk theory. Our simulations show how BH spin, magnetic field, and jet
mass-loading affect the radiative and jet efficiencies of
super-Eddington accretion.

\end{abstract}

\begin{keywords}
accretion, black hole physics, (magnetohydrodynamics) MHD, radiation
\end{keywords}

\section{Introduction}
\label{sec_intro}
\newcommand{\MBH}{{M}}
\newcommand{\MBHO}{{M_i}}
\newcommand{\Mdot}{{\dot{M}}}
\newcommand{\Mdotedd}{{\dot{M}_{\rm Edd}}}

Black hole (BH) accretion drives a broad range of phenomena, including
quasars, active galactic nuclei, BH X-ray binaries, some gamma-ray
bursts, and tidal disruption events (TDEs).  The disk's gravitational
potential energy and BH spin energy are converted into radiation,
winds, and relativistic jets via magnetic stresses generated by the
magneto-rotational instability (MRI) \citep{1998RvMP...70....1B},
magnetic field threading the disk \citep{bp82}, or magnetic field
threading the BH \citep{bz77}.  Magnetic torques are maximized when
magnetic flux advects inward and piles-up until magnetic stresses
balance incoming gas forces -- the so-called magnetically-arrested
disk (MAD) state \citep{2003PASJ...55L..69N}.  Simulations of
non-radiative MADs show up to $\sim 300\%$ total efficiency
\citep{tnm11,2012MNRAS.423.3083M}, corresponding to $3$ times the
energy out as going into the BH.  A fundamental question is: What form
does this energy take at large distances and what fraction is
radiative?

The radiative efficiency of BH systems is expected to be dependent
upon the accretion rate $\Mdot$, which controls the density, disk
thickness, and dynamical importance of radiation (from gas-pressure
dominated at low $\Mdot$ to radiation-pressure dominated at high
$\Mdot$).  To scale $\Mdot c^2$ or the radiative luminosity $L$, for a
black hole mass $\MBH$, speed of light $c$, gravitational constant $G$
(giving gravitational radius $r_g\equiv GM/c^2$), and Thomson electron
scattering opacity $\kappa_{\rm es}$, one can use the Eddington
luminosity
\begin{equation}\label{Ledddef}
L_{\rm Edd}=\frac{4\pi G \MBH c}{\kappa_{\rm es}} \approx 1.3\times 10^{46} \frac{\MBH}{10^8M_{\odot}}{\erg~{\rm s}^{-1}} .
\end{equation}
One can also choose to normalize $\Mdot$ by
$\Mdotedd = (1/\eta_{\rm NT})L_{\rm Edd}/c^2$, where $\eta_{\rm NT}$
is the nominal accretion efficiency for the Novikov-Thorne thin disk
solution \citep{1973blho.conf..343N} (commonly, a fixed $\eta_{\rm
  NT}=0.1$ is used, but we include the spin dependence).

For luminosities $L\gtrsim 0.3 L_{\rm Edd}$, the accretion flow is
expected to become geometrically thick and optically thick, and in
this regime the photons can remain trapped within the flow as in the
``slim disk'' model (which includes no magnetic field) \citep{abr88}.
The super-Eddington accretion regime where $\dot{M}\gtrsim
\dot{M}_{\rm Edd}$ or $L\gtrsim L_{\rm Edd}$ may help explain
ultra-luminous X-ray sources as highly super-Eddington stellar-mass
BHs \citep{wata01,miller04}.  Also, some black hole X-ray binaries can
have $L\gtrsim L_{\rm Edd}$ (e.g., SS433,
\citealt{margon79,2010PASJ...62L..43T}; GRS1915+105, \citealt*{fb04}),
while tidal disruption events seem to require $\dot{M}\gg \dot{M}_{\rm
  Edd}$ \citep{2011Sci...333..203B}.

In this paper, we seek to test whether super-Eddington MADs around
spinning BHs are radiatively efficient. We use the general
relativistic radiation magnetohydrodynamic (GRRMHD) code HARMRAD,
which uses the M1 closure for radiation \citep{2014MNRAS.441.3177M}.
The slim disk model and recent radiative GRMHD simulations of non-MAD
(i.e. lower magnetic flux than MAD on the BH and in the disk)
super-Eddington flows are radiatively inefficient with $L\lesssim
L_{\rm Edd}$ even for $\dot{M}\sim 100\dot{M}_{\rm Edd}$
\citep{2014MNRAS.439..503S,2014MNRAS.441.3177M}, far below the thin
disk Novikov-Thorne (NT) efficiency (e.g. about $6\%$ efficient for BH
spin of $a/M=0$ to $12\%$ efficient for $a/M=0.8$).  However, MADs
(not yet studied with radiation) can maximize the efficiency and may
lead to a radiatively efficient state.  This is also plausible for
rapidly spinning BHs because MADs are then compressed into a thin disk
\citep{2012MNRAS.423.3083M}, which can lose radiation more rapidly.

We discuss the physical and numerical setup in \S\ref{sec:fiducial}.
Results and discussions are presented in \S\ref{sec:results}.  We
summarize in \S\ref{sec:summary}.

\section{Fully 3D GR Radiative MHD MAD Model}
\label{sec:fiducial}

This model with black hole mass of $\MBH=10M_{\odot}$ and
dimensionless spin $a/M=0.8$ has nominal thin disk efficiency
$\eta_{\rm NT} \approx 12.2\%$, so that $\Mdotedd \approx 1.2\times
10^{19}$g/s (see Eq.~\ref{Ledddef}).  While we choose a specific BH
mass, the flow is strongly electron scattering dominated and so the
results might apply roughly equally to all BH masses.

\subsection{Opacities}

We assume solar abundances (mass fractions of Hydrogen, Helium, and
``metals'', respectively, $X=0.7$, $Y=0.28$, $Z=0.02$) giving
electron fraction $Y_e=(1+X)/2$ and mean molecular weight
$\bar{\mu}\approx 0.62$, entering gas entropy, pressure, and
temperature $T_g$ [Kelvin].

We use a frequency ($\nu$) mean of the opacity $\alpha_\nu$ to get an
absorption-mean absorption opacity (units of [${\rm cm}^2/{\rm g}$])
of $\kappa = (\int_\nu d\nu \alpha_\nu J_\nu)/(\int_\nu d\nu J_\nu)$,
where the absorbed radiation $J_\nu$ is assumed to be a Planck
distribution at a radiation temperature of $T_r = (\hat{E}/a_{\rm
  rad})^{1/4}$ [Kelvin], where $\hat{E} = u^\mu u^\nu R_{\mu\nu}$,
$u^\mu$ is fluid 4-velocity, $R$ is radiation stress-energy tensor,
and $a_{\rm rad}$ is the radiation constant.

The electron scattering opacity  is
\begin{equation}
\kappa_{\rm es} \approx 0.2 (1+X) \kappa_{\rm kn} ,
\end{equation}
where the Klein-Nishina correction for thermal electrons is
$\kappa_{\rm kn} \approx (1 + (T_g/(4.5\times 10^8))^{0.86})^{-1}$
\citep{1976ApJ...210..440B}.

The absorption-mean energy absorption opacity is
\begin{equation}
\kappa_{\rm abs} \approx \left(\frac{1}{\kappa_{\rm m} + \kappa_{H^-}} + \frac{1}{\kappa_{\rm Chianti} + \kappa_{\rm bf} + \kappa_{\rm ff}}\right)^{-1} ,
\end{equation}
which bridges between different temperature regimes.  The molecular
opacity is $\kappa_{\rm m} \approx 0.1Z$, and the $H^-$ opacity is
$\kappa_{H^-} \approx 1.1\times 10^{-25}Z^{0.5} \rho^{0.5} T_g^{7.7}$.
Rest-mass density, $\rho$, is in cgs units.  The Chianti opacity
accounts for bound-free at slightly lower temperatures and is given by
$\kappa_{\rm Chianti} \sim 4.0\times 10^{34} \rho (Z/Z_{\rm solar})
Y_e T_g^{-1.7} T_r^{-3.0}$ (this accounts for the assumed $Z=Z_{\rm
  solar} = 0.02$ for figure 34.1 in \citet{2011piim.book.....D}, most
applicable for baryon densities of $n_b\sim 1{\rm cm}^{-3}$).  The
bound-free opacity is
\begin{equation}
\kappa_{\rm bf} \approx 3\times 10^{25} Z (1+X+0.75Y) \rho T_g^{-0.5} T_r^{-3.0} \ln\left(1 + 1.6 (T_r/T_g)\right) ,
\end{equation}
\citep{1986rpa..book.....R}, where the $1+X+0.75Y$ term is roughly accurate
near solar abundances, and the $\ln()$ term comes from fitting the
absorption-mean integral.  The free-free opacity is
\begin{eqnarray}
\kappa_{\rm ff} &\approx& 4\times 10^{22} (1+X)(1-Z) \rho T_g^{-0.5} T_r^{-3.0} \ln\left(1 + 1.6 (T_r/T_g)\right)\nonumber \\
&\times& \left(1 + 4.4\times 10^{-10}T_g\right) ,
\end{eqnarray}
for thermal electrons and no pairs (see Eq.5.25 in
\citealt{1986rpa..book.....R} and \citealt{1991pav..book.....S}).
Thermal energy Comptonization is included as in
\citet{2015MNRAS.447...49S}.  The mean emission opacity $\kappa_{\rm
  emit}$ is the same as $\kappa_{\rm abs}$ but letting $T_r\to T_g$,
such that Kirchoff's law gives an energy density emission rate of
$\lambda = c \rhorest \kappa_{\rm emit} a_{\rm rad} T_g^4$.  The total
opacity is $\kappa_{\rm tot} = \kappa_{\rm es} + \kappa_{\rm abs}$.
The low-temperature opacities avoid unphysical opacity divergences
during the simulation.

\subsection{Initial Conditions}

The initial disk is Keplerian with a rest-mass density that is
Gaussian in angle with a height-to-radius ratio of $H/R\approx 0.2$
and radially follows a power-law of $\rhorest\propto r^{-0.6}$.  The
solution near and inside the inner-most stable circular orbit (ISCO)
is not an equilibrium, so near the ISCO the solution is tapered to a
smaller density ($\rhorest\to \rhorest (r/15)^7$, within $r=15r_g$)
and a smaller thickness ($H/R\to 0.2 (r/10)^{0.5}$, within $r=10r_g$
-- based upon a low-resolution simulation).  The total internal energy
density $u_{\rm tot}$ is estimated from vertical equilibrium of
$H/R\approx c_s/v_K$ for sound speed $c_s\approx \sqrt{\Gamma_{\rm
    tot} P_{\rm tot}/\rhorest}$ with $\Gamma_{\rm tot}\approx 4/3$ and
Keplerian speed $v_K\approx (r/r_g)/((r/r_g)^{3/2}+a/M)$.  The total
ideal pressure $P_{\rm tot} = (\Gamma_{\rm tot}-1) u_{\rm tot}$ is
randomly perturbed by $10\%$ to seed the MRI.  The disk gas has
$\Gamma_{\rm gas}=5/3$.  The disk has an atmosphere with
$\rhorest=10^{-5} (r/r_g)^{-1.1}$ and gas internal energy density
$\ug=10^{-6} (r/r_g)^{-5/2}$.  The disk's radiation energy density and
flux are set by local thermal equilibrium (LTE) and flux-limited
diffusion \citep{2014MNRAS.441.3177M} with a negligible radiation
atmosphere.

We do not use polish donuts \citep{abramowiczetal78} or equilibrium
tori as initial conditions.  The outer parts of tori have an extended
column of gas at high latitudes that contributes significantly to
spurious luminosity (see section 6.8 in
\citealt{2014MNRAS.441.3177M}), trapping of the disk wind and
radiation, and artificial (instead of the self-consistent wind)
collimation of the jet.

The initial magnetic field is large-scale and poloidal.  For
$r<300r_g$, the coordinate basis $\phi$-component of the vector
potential is
\begin{equation}
A_\phi = {\rm MAX}(r^\nu 10^{40} - 0.02,0) (\sin\theta)^{1+h} ,
\end{equation}
with $\nu=1$ and $h=4$.  For $r\ge r_0 = 300r_g$, the field
transitions to monpolar using $A_\phi = {\rm MAX}(r_0^\nu 10^{40} -
0.02,0) (\sin\theta)^{1+h(r_0/r)}$.  The field is normalized with
$\sim 1$ MRI wavelength per half-height $H$ giving a ratio of average
gas+radiation pressure to average magnetic pressure of $\beta\approx
24$ for $r<100r_g$.

\subsection{Numerical Grid and Density Floors}
\label{sec:numsetupgrid}

The numerical grid mapping equations and boundary conditions used here
are identical to that given in \citet{2012MNRAS.423.3083M}.  The
radial grid of $N_r=256$ cells spans from $R_{\rm in}\approx
0.688r_{\rm H}$ (horizon radius $r_{\rm H}$) to $R_{\rm out}=10^5r_g$
with cell size increasing exponentially till $r\sim 500r_g$ and then
even faster.  Radial boundaries use absorbing conditions.  The
$\theta$-grid of $N_\theta=128$ cells spans from $0$ to $\pi$ with
mapping given in \citet{2012MNRAS.423.3083M} but with $n_{\rm
  jet}=0.7$ to follow the jet, and other coefficients are slightly
tuned so that the grid aspect ratio at $r\sim 30r_g$ is $1:2:3$.
Approaching the horizon (where the disk thins due to magnetic
compression), the grid is tuned to follow the compressed disk, such
that on the horizon there are $20$ points across a half-height of the
actual final disk with thickness of order $H/R\sim 0.1$.  The
Poynting-dominated polar jet contains, respectively, about $80,60,110$
$\theta$ grid cells near the horizon, $r\sim 20r_g$, and $r\gtrsim
500r_g$.  This gives sufficient resolution of the Poynting-dominated
jet.  The $\phi$-grid of $N_\phi=64$ cells spans uniformly from $0$ to
$2\pi$ with periodic boundary conditions.

As in our other papers
\citep{2012MNRAS.423.3083M,2013Sci...339...49M,2014MNRAS.441.3177M},
we test the so-called convergence quality factors for the MRI in the
$\theta$ and $\phi$ directions ($Q_{\theta,\rm MRI}$ and $Q_{\phi,\rm
  MRI}$) and turbulence ($Q_{nlm,\rm cor}$) measuring, respectively,
the number of grid cells per MRI wavelength in the $\theta$ and $\phi$
directions and the number of grid cells per correlation length in the
radial, $\theta$ and $\phi$ directions.  At late times, our simulation
has $Q_{\theta,\rm MRI}\sim 340$, $Q_{\phi,\rm MRI}\sim 45$, and
rest-mass and magnetic energy densities have $Q_{nlm,\rm cor}\sim
25,20,6$ at $r\sim 8r_g$, indicating good $r,\theta$ resolution and
marginal $\phi$ resolution.  We also measure the thickness of the disk
per unit MRI wavelength, $S_d$, where $S_d<0.5$ is where the MRI is
suppressed.  Initially $S_d\sim 0.6$ at all radii, while the
time-averaged flow has $S_d\sim 0.1$ out to $r\sim 60r_g$.

The rest-mass and internal energy densities are driven to zero near
the BH within the jet and near the axis, so we use numerical
ceilings of $b^2/\rhorest = 300$, $b^2/\ug=10^9$, and
$\ug/\rhorest=10^{10}$.  The value of $b^2/\rhorest$ is at the code's
robustness limit for the chosen resolution.

\subsection{Diagnostics}
\label{sec:diagnostics}

The disk's geometric half-angular thickness ($H$) per radius ($R$) is
\begin{equation}\label{thicknesseq}
\frac{H}{R}(r,\phi) \equiv \frac{H_0}{R} + \frac{\left(\int_\theta \rho(\theta-\theta_0)^n dA_{\theta\phi}  \right)^{1/n}}{\left(\int_\theta \rho dA_{\theta\phi} \right)^{1/n}} ,
\end{equation}
with $n=2$, $H_0/R=0$, and surface differential $dA_{\theta\phi}$.
One computes $\theta_0$ like $H/R$, but let $n=1$ and
$\{\theta_0,H_0/R\}=\pi/2$.

The mass accretion rate and energy efficiency are, respectively,
\begin{eqnarray}\label{Dotsmej}
\Mdot  &=&  \left|\int\rhorest u^r dA_{\theta\phi} \right| , \\
\eff   &=& -\frac{\int (T^r_t+\rhorest u^r+R^r_t) dA_{\theta\phi}}{[\Mdot]_H} ,
\end{eqnarray}
where $T$ is the gas stress-energy tensor and $[\Mdot]_H$ is the
time-averaged $\dot{M}$ on the horizon.  The specific angular momentum
accreted is $\jmath = (\int (T^r_\phi+R^r_\phi)
dA_{\theta\phi})/[\Mdot]_H$.  Both $\eff$ and $\jmath$ are composed of
free particle (PAKE=kinetic+gravitational), thermal (EN),
electromagnetic (EM), and radiation (RAD) components.  The jet is
defined as PAKE+EN+EM, in locations where magnetic energy density
exceeds rest-mass energy density. The wind is defined as PAKE+EN,
located outside the jet. The wind can also contain untapped EM and RAD
components.  The dimensionless magnetic flux is
\begin{equation}\label{equpsilon}
\Upsilon \approx 0.7\frac{\int dA_{\theta\phi} 0.5|\Bvec^r|}{\sqrt{[\Mdot]_H}} ,
\end{equation}
for field $\Bvec$ in Heaviside-Lorentz units
\citep{2014MNRAS.441.3177M}.

To obtain the optical depth, at each instant in time we compute
\begin{equation}\label{tau}
\tau\approx \int \rhorest \kappa_{\rm tot} dl .
\end{equation}
For the radial direction, $dl=- f_\gamma dr$, $f_\gamma \approx u^t (1
- (v/c)\cos\theta)$, $(v/c)\approx 1-1/(u^t)^2$ (as valid at large
radii), $\theta=0$, and the integral is from $r_0=3000$ (being some
radius beyond which only transient material would contribute to the
optical depth, but a radius the disk wind has reached) to $r$ to
obtain $\tau_r(r)$.  For the angular direction, $dl=f_\gamma r
d\theta$, $\theta=\pi/2$, and the integral is from each polar axis
toward the equator to obtain $\tau_\theta(\theta)$.  The flow's ``true
radiative photosphere'' is defined as when $\tau_r=1$, a conservative
upper limit to the radius of the photosphere for an observer, because
radiation can escape by tracking with relativistic low-density parts
of the jet.

The radiative luminosity is computed at each instant as
\begin{equation}
L = -\int dA_{\theta\phi} R^r_t ,
\end{equation}
which is usually measured at $r=400r_g$, where we only include those
angles where the gas has $\tau_r(r)<1$.

\section{Results and Discussion}\label{sec:results}

\begin{figure}
\centering
\includegraphics[width=3.2in,clip]{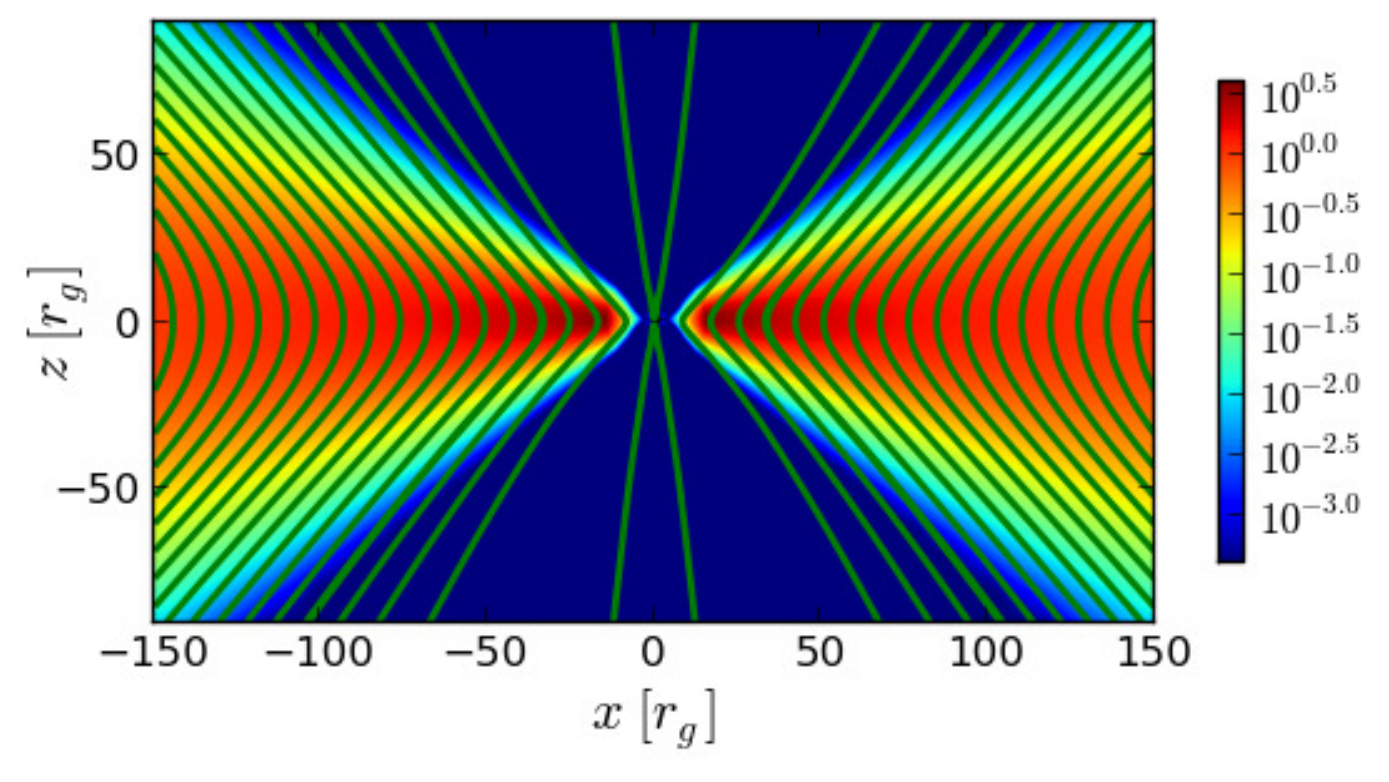}
\caption{The initial ($t=0$) state consists of a weakly magnetized
  radially-extended vertically-Gaussian disk with $H/R\sim 0.2$ around
  a spinning ($a/M=0.8$) BH. Rest-mass density is shown as color with
  legend, while green lines show magnetic field lines.  Density values
  are made dimensionless using $\dot{M}_{\rm Edd}$, length $r_g$, and
  time $r_g/c$.  The initial disk is threaded by weak (but ordered)
  magnetic flux, capable of accumulating onto the BH and leading to
  the MAD state.}
\label{initial3plot}
\end{figure}

\begin{figure}
\centering
\includegraphics[width=3.2in,clip]{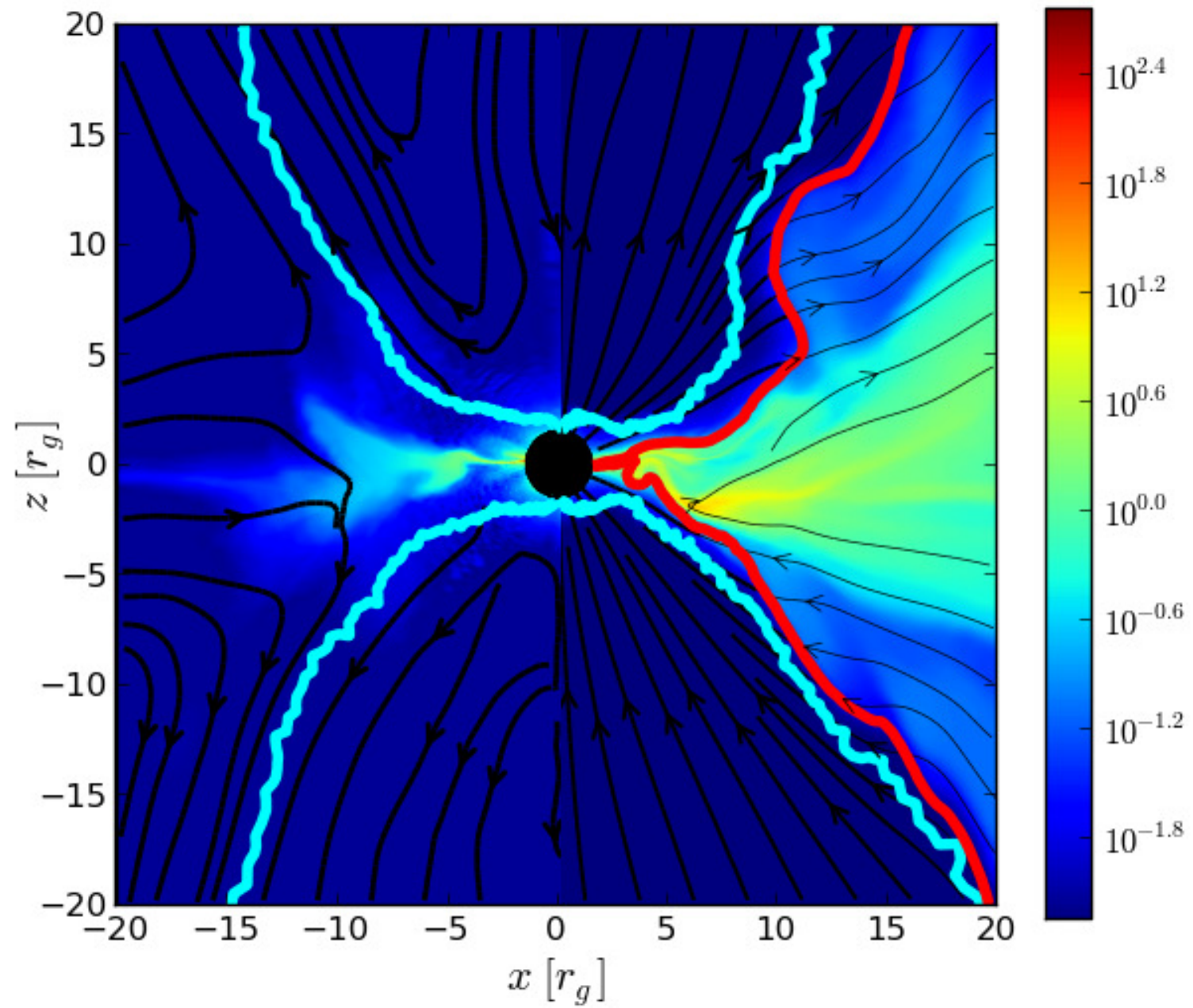}
\caption{The evolved ($t=31,200r_g/c$) super-Eddington MAD state, with
  the figure split at $x=0$ as two panels ($x<0$ on left and $x>0$ on
  right).  Left panel shows radiation-frame radiation energy density
  (color, with legend), radiation velocity lines (black, fixed line
  width), and optical depth of unity away from each polar axis,
  $\tau_\theta=1$ (cyan lines).  Right panel shows fluid-frame
  rest-mass density (color, same legend), magnetic field lines (black,
  thicker lines for more magnetized gas), where magnetic energy is
  equal to rest-mass energy density (red lines), and same optical
  depth (cyan lines).  The MAD state reaches a quasi-steady state out
  to $r\sim 20r_g$.  Radiation advects inward within the equatorial
  disk, outward through the low-density jet channel, and outward along
  with the optically thick wind.}
\label{evolvedmovie}
\end{figure}

Fig.~\ref{initial3plot} and Fig.~\ref{evolvedmovie} show the initial
and final state of the accretion flow.  The initial disk is threaded
by a weak large-scale poloidal magnetic field around a spinning black
hole with $a/M=0.8$.  Rotation amplifies the magnetic field via the
MRI leading to accretion of mass, energy, angular momentum, and
magnetic flux.  The magnetic flux accumulates and eventually forms a
quasi-steady super-Eddington ($\Mdot\sim 400L_{\rm Edd}/c^2\sim
50\Mdotedd$) state that is a MAD (where the MRI is suppressed) out to
$r\sim 60r_g$ after the model is evolved for a time $31,200r_g/c$.
The magnetically-compressed radiatively efficient thin disk near the
BH is exposed by the jet channel, and the magnetized wind carries a
significant fraction of radiation away from the disk.

\begin{figure}
\centering
\includegraphics[width=3.2in,clip]{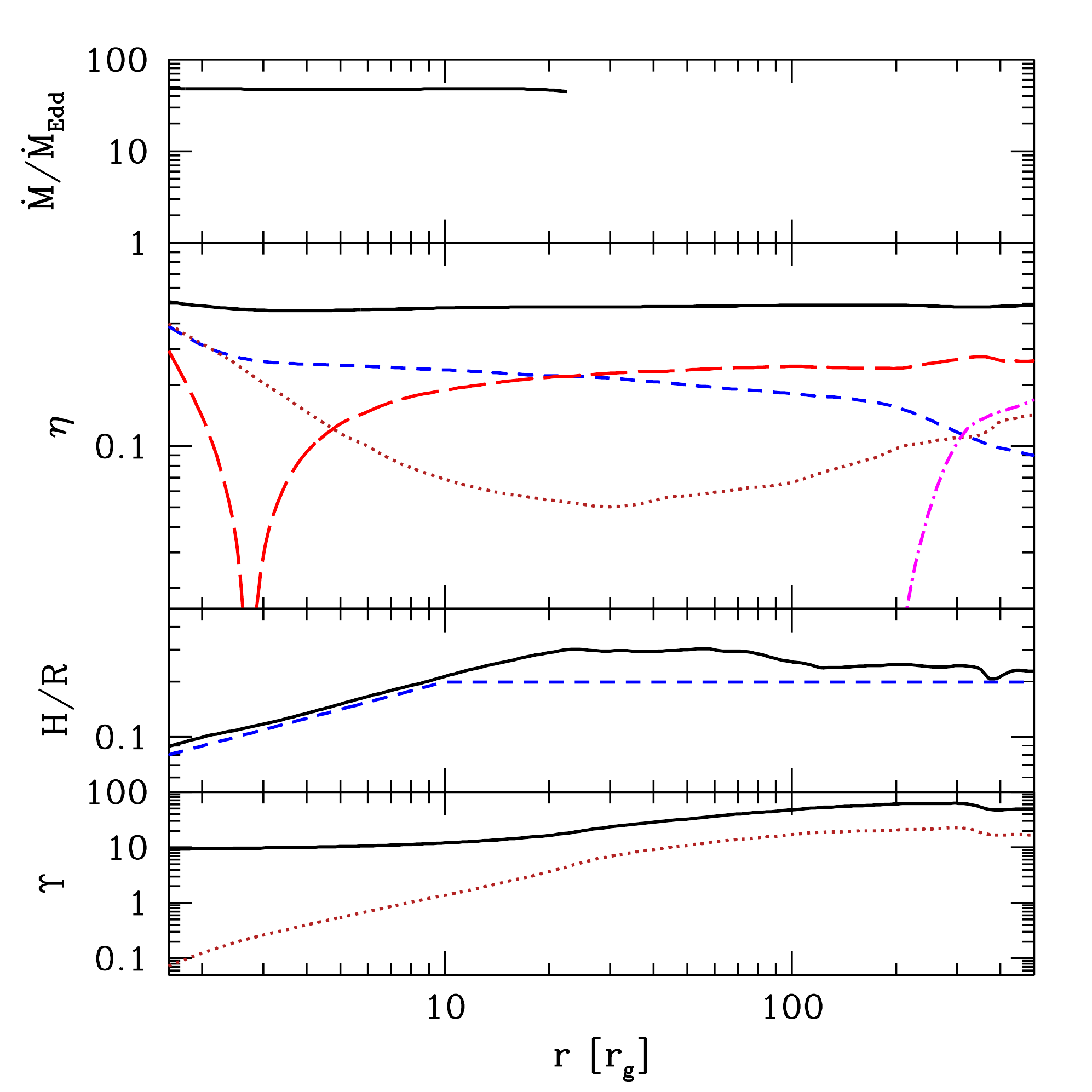}
\caption{The time-averaged angle-integrated mass flux, efficiency,
  disk thickness, and magnetic flux.  From top to bottom, panels are:
  Total mass accretion rate ($\Mdot$) per unit $\dot{M}_{\rm Edd}$ out
  to where there is inflow and thermal equilibrium ($r\sim 20r_g$,
  beyond which the line is truncated), energy efficiency $\eta$
  (total: solid black line, electromagnetic: blue dashed, matter
  without rest-mass (i.e. kinetic+gravitational+thermal): dark red
  dotted, radiation (negative within $3r_g$): red long-dashed,
  radiative luminosity $L$: magenta dot-dashed), disk thickness $H/R$
  (evolved state: black solid, initial state: blue dashed), and
  magnetic flux (total on BH and threading disk: black solid, only
  threading disk: dark red dotted). The disk is magnetically
  compressed near the BH, and the thin disk generates a significant
  radiative flux that effectively releases starting at $r\sim 300r_g$.
  The initial disk with $H/R\approx 0.2$ has puffed up to $H/R\approx
  0.3$ from $r\sim 20$--$70r_g$ after evolving for several thermal
  times.  By $r\sim 400r_g$, the luminosity reaches $L\sim 50L_{\rm
    Edd}$.}
\label{figure3}
\end{figure}

Fig.~\ref{figure3} shows the mass accretion rate, efficiencies, disk
thickness, and magnetic flux as time-averaged from $t=30,000r_g/c$ to
$31,200r_g/c$. The disk is in inflow and thermal equilibrium out to
$r\sim 20r_g$ with constant fluxes of mass, energy, and specific
angular momentum vs. radius.  The total (kinetic + gravitational +
electromagnetic + radiative) efficiency $\eta\approx 50\%$. As in
non-radiative MAD models \citep{tnm11,2012MNRAS.423.3083M}, the
efficiency is beyond the Novikov-Thorne value of $\eta_{\rm NT}\approx
12\%$.

The magnetic field threading the BH is strong, with $\Upsilon\approx
10$ at late time and $\Upsilon\approx 8$ on average with $B_z \propto
r^{-5/4}$ through the disk.  The magnetic field threading the BH and
disk leads to a magnetized wind that carries a significant amount of
radiation away from the disk and helps to avoid the classical photon
trapping effect of slim disks \citep{abr88}. The magnetized wind at
$r\sim 400r_g$ has a matter efficiency of about $15\%$, but the
optically thick wind also contains about $15\%$ trapped radiation
energy and $10\%$ electromagnetic energy that could be tapped for
acceleration or heating. By $r\sim 400r_g$, the wind carries most of
its radiation within a half-opening angle of $30^\circ$ around the
polar axes.

The Blandford-Znajek (BZ) effect leads to an electromagnetic, BZ,
efficiency $\eta\sim 35\%$ on the BH.  Some of that energy forms a jet
with $\eta\sim 10\%$ at $r=50r_g$, but some of that energy is absorbed
by the wind, leading to a jet having only $\eta\sim 5\%$ by
$r=400r_g$.

Radiative-loading of magnetic field lines threading the BH leads to a
lower-than-optimal BZ efficiency (relatedly, see
\citealt{2004MNRAS.347..587B,2015PASJ..tmp..160T}) once the MAD state
builds-up.  The inertial loading of magnetic field lines is due to a
total energy density $\rho_{\rm tot}=\rhorest + \ug + \hat{E}$ once
$\tau\gtrsim 1$ caused by $\rhorest$ keeping up with the MAD's higher
$b^2$ when the numerical ceiling of $b^2/\rhorest=300$ is enforced.
The optically thick radiation slows the jet magnetic field line
rotation rate down by order unity compared to the optimal BZ value,
because $b^2/\rho_{\rm tot}\sim 1$ in the funnel (even if
$b^2/\rhorest\gg 1$ there) as determined by the rough condition that
the jet and disk have similar $b^2$, yet $\Upsilon\gg 1$ implies that
$b^2\sim \rhorest$ in the disk, and radiative energy density
$\hat{E}\sim \rhorest$ (from a thermal estimate of $H/R\sim
c_s/v_K\sim 0.3$).  A restarted simulation with an exponentially
reduced (as $b^2/\rhorest$ approached its ceiling) opacity has a BZ
efficiency of $\eta\sim 100\%$ on the BH, as consistent with
$\Upsilon\approx 8$ \citep{tnm11}.  A restarted simulation with a
lower ceiling $b^2/\rhorest=100$ led to a $20\%$ BZ efficiency.  So,
mass-loading and opacity physics in the funnel are important to
BZ-driven jets and winds for super-Eddington flows where radiation
crosses magnetic field lines.

\begin{figure}
\centering
\includegraphics[width=3.1in,clip]{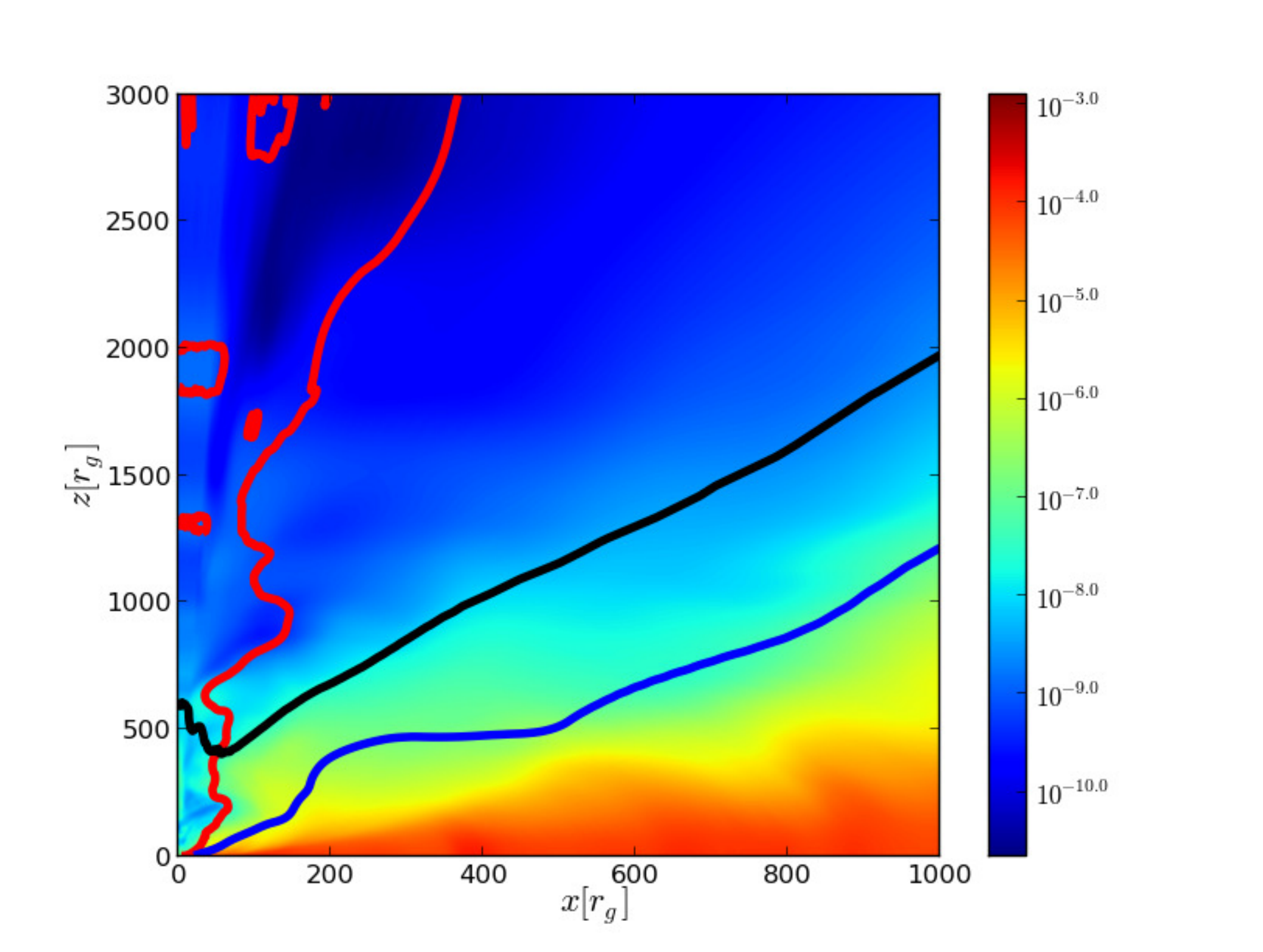}
\caption{The evolved ($t=31,200r_g/c$) super-Eddington MAD state at
  large radii showing the jet, wind, and photosphere. Shows
  fluid-frame rest-mass density (color, with legend), where magnetic
  energy equal to rest-mass energy density (red lines), radiative
  photosphere at $\tau_r=1$ (black line), and the wide-angle wind
  component's magnetic field line, within which (toward the polar
  axis) the outflow has achieved more than a single flow time along
  the line extending $\sim 1000r_g$ (blue line).  Quantities are
  $\phi$-averaged.  The jet extends to $r\sim 30,000r_g$ with Lorentz
  factor $\gamma\sim 5$ by $r\sim 10^3r_g$.  The wind extends to
  $r\sim 3000r_g$, with most of the wind's trapped radiation having an
  opening angle of $30^\circ$ at $r\sim 400r_g$. The disk at $R\gg
  100r_g$ has not evolved much.  Radiation within $r\lesssim 400r_g$
  can more freely stream within the jet's low-density channel. By
  $t\sim 30,000r_g/c$, at $r\sim 400r_g$ within $15^\circ$ around the
  polar axis, the outflow become conical and optically thin.}
\label{largeradius}
\end{figure}

Fig.~\ref{largeradius} shows the jet, wind, and ``true'' photosphere
at large radii.  The jet-wind boundary oscillates in angle to order
unity, causing the wind to absorb some of the jet and radiation
energy.  We have resolved the true photosphere at $r\sim 400r_g$, and
the radiation escaping to an observer has a high radiative efficiency
of $\eta\approx 15\%$ (comparable with the NT value of $\eta\sim
12\%$), corresponding to luminosity $L\sim 50L_{\rm Edd}$.  Most of
this radiation is within a half-opening angle of $15^\circ$ around the
polar axes.  The luminosity in this GRRMHD MAD simulation is much
higher than predicted by the slim disk solution at $\eta\sim 2\%$ or
seen in other viscous non-MAD or non-MAD MHD simulations at
$\eta\lesssim 1\%$
\citep{Ohsuga+05,om11,2014MNRAS.439..503S,2014MNRAS.441.3177M}.  Far
beyond the photosphere, radiation might be absorbed
\citep{2015arXiv150300654S} or some might leak out of the wind
containing $\eta\approx 15\%$ in trapped radiation, but our grid is
too unresolved to accurately track energy conversion for $r>600r_g$.
The radiation ultimately released is high in our simulation because 1)
the disk near the BH is forced to become thin; 2) a magnetized wind
pulls radiation-filled material off the disk; 3) the jet drills a
channel and pushes back the opaque disk wind; and 4) the jet becomes
conical at large radii allowing free-streaming of radiation.  These
lead to an effective diffusion timescale shorter than in the slim disk
model.

We also performed otherwise identical simulations that are MAD with
$a/M=0$ as well as non-MAD simulations using a toroidal field in the
initial disk (with $\beta\sim 20$ for all radii at the equator).  The
non-spinning MAD never produced a jet and the radiation remains
trapped by the broad disk wind.  Similarly, the toroidal field models
with $a/M=0$ and $a/M=0.8$ show no low-density funnel region.  Similar
to \citet{2015arXiv150300654S}, our $a/M=0$ MAD model with
$\dot{M}\sim 20\dot{M}_{\rm Edd}$ has a wind with NT-level efficiency
of order $5\%$ and negligible radiation flux at large radii by $r\sim
400r_g$ within the photosphere, because all radiation energy flux
converted to kinetic energy flux.  Our toroidal field models with
$a/M=0$ and $a/M=0.8$ have $\dot{M}\sim \dot{M}_{\rm Edd}$ with a wind
efficiency of $2\%$ and $4\%$, respectively, and a radiative
efficiency of about $1\%$.  So the formation of a low-density jet
channel by a spinning BH, with enough magnetic flux to launch a jet
and strong wind, helps super-Eddington accretion become radiatively
efficient.

Convergence testing was performed by restarting each
previously-mentioned model at a resolution twice lower in each
dimension.  The restarted simulation starts at half-way through the
higher resolution simulation, and then the two resolutions are
compared for the latter half of the simulations.  We find that the
total, BH, jet and radiative efficiencies, and $\Upsilon$ agree to
within $30\%$ or smaller error.  Our convergence quality factors for
the MRI and turbulence also suggest the simulations are converged.

A non-relativistic MHD simulation with $a/M=0$ by
\citet{2014ApJ...796..106J} measured a NT-level radiative efficiency
of about $5\%$ that they attributed to MRI-driven magnetic buoyancy
vertical transport of radiation.  Their simulation had a small
vertical extent $\pm 60r_g$, so $\tau$ computed from Eq.~\ref{tau}
does not include the extended wind.  Over extended distances, a
significant portion of radiation energy flux can convert into wind
energy flux \citep{2015arXiv150300654S}.  In our $a/M=0.8$ MAD
simulation, we attribute the high radiative efficiency to magnetic
compression of the disk into a thin MAD, magnetized wind advection of
radiation away from the disk, and the formation of a low-density jet
channel through which radiation can more freely stream.  Our
simulation boundary is at $r\sim 10^5r_g$, with accurate energy
conversion out to $r\sim 600r_g$.  Our photosphere at $r\sim 400r_g$
accounts for wind material that reached $r\gtrsim 3000r_g$.  If we
ignore the extended wind, for the $a/M=0.8$ MAD model we would
miscalculate the radiative efficiency to be $\eta\sim 30\%$ -- an
overestimate by a factor of two.  For the $a/M=0$ MAD model, measuring
at $r\sim 60r_g$ gives $\eta\sim 5\%$ -- an overestimate from
$\eta\sim 0\%$ (caused the wind absorbing radiation energy).

\section{Summary}
\label{sec:summary}

We have performed fully 3D simulations of super-Eddington accretion,
including a simulation with $\Mdot\sim 400L_{\rm Edd}/c^2\sim
50\Mdotedd$ onto a rotating black hole with $a/M=0.8$.  Sufficient
magnetic flux was distributed throughout the disk that the BH and disk
reached MAD levels, where magnetic forces pushing out balance gas
forces pushing in.  The MAD state enabled the super-Eddington
accretion flow to reach its maximum efficiency, with the total
efficiency measured to be about $50\%$ (higher for higher BH spins).
Importantly, the system has a high radiative efficiency of about
$15\%$ (luminosity $L\sim 50L_{\rm Edd}$) beyond the resolved
photosphere at $r\sim 400r_g$.  This occurs because the magnetized
wind carries radiation away from the disk. Also, the magnetized jet
creates a low-density channel for radiation to more freely stream and
pushes away the more opaque wind.  These effects increase the
radiative flux escaping the disk and diminish conversion of radiation
energy flux into kinetic energy flux of the wind.  This mechanism
allows high radiative efficiencies from super-Eddington accretion
systems.

As applied to jetted TDEs \citep{2014MNRAS.437.2744T}, our
super-Eddington MAD simulations show how the jet and radiative
efficiencies depend not only upon BH spin and magnetic flux but also
upon jet mass-loading physics that leads to sub-optimal BZ
efficiencies via dragging of BH field lines by optically thick
radiation.  Also, while the wind absorbs some jet energy, the jet
channel exposes the (otherwise obscured) inner hot X-ray emitting
disk.

\section*{Acknowledgments}

We thank Ramesh Narayan, Alexander Tchekhovskoy, Yan-Fei Jiang, and
Aleksander Sadowski for discussions and acknowledge NASA/NSF/TCAN
(NNX14AB46G), NSF/XSEDE/TACC (TG-PHY120005), and
NASA/Pleiades (SMD-14-5451).

%\section*{SUPPORTING INFORMATION}

%Additional Supporting Information may be found in the online version
%of this article: 2 movie files and captions.

\bibliographystyle{mnras}
{\small
\bibliography{mybibnew}

}

\label{lastpage}
\end{document}